\documentclass[10pt,conference]{IEEEtran}
\usepackage[a4paper,left=1.2cm,right=1.2cm,top=1.8cm,bottom=3.7cm]{geometry}
\usepackage{cite}
\usepackage{amsmath,amssymb,amsfonts}
\usepackage{algorithmic}
\usepackage[ruled,vlined,linesnumbered]{algorithm2e}
\SetAlgorithmName{Algorithm}{Algorithm}{List of Algorithms}
\usepackage{graphicx}
\usepackage{stmaryrd}
\usepackage{xcolor}
\usepackage{array}
\usepackage{commath}
\usepackage{sidecap}
\usepackage{stfloats}
\usepackage{float}
\usepackage{tabularx, boldline}
\usepackage{rotating,booktabs,multirow}
\usepackage{mathtools}
\usepackage{flexisym}
\usepackage{mathtools}
\usepackage{amsmath}
\usepackage{graphicx}
\usepackage{tabularx, boldline}
\usepackage{rotating,booktabs,multirow}
\usepackage{enumerate}
\usepackage{etoolbox}
\usepackage[]{nomencl}   
    \makenomenclature

\def\R{\mathbb{R}} 
\def\E{\mathbb{E}} 
\newcommand{\mkv}{-\!\!\!\!\minuso\!\!\!\!-}

\ifCLASSINFOpdf
\else
\fi

\begin{document}

\title{On the Impact of Side Information on Smart Meter Privacy-Preserving Methods}

\author{\IEEEauthorblockN{Mohammadhadi Shateri}
\IEEEauthorblockA{\textit{McGill University} \\  Montreal, Canada}
\and
\IEEEauthorblockN{Francisco Messina}
\IEEEauthorblockA{\textit{McGill University} \\  Montreal, Canada}
\and
\IEEEauthorblockN{Pablo Piantanida}
\IEEEauthorblockA{CentraleSup\'elec-CNRS-Universit\'e Paris-Saclay\\
 Gif-sur-Yvette, France}
\and
\IEEEauthorblockN{Fabrice Labeau}
\IEEEauthorblockA{\textit{McGill University} \\  Montreal, Canada}}

\maketitle

\begin{abstract}
Smart meters (SMs) can pose privacy threats for consumers, an issue that has received significant attention in recent years. This paper studies the impact of Side Information (SI) on the performance of possible attacks to real-time privacy-preserving algorithms for SMs. In particular, we consider a deep adversarial learning framework, in which the desired releaser,  which is a Recurrent Neural Network (RNN), is trained by fighting against an adversary network until convergence. To define the objective for training, two different approaches are considered: the Causal Adversarial Learning (CAL) and the Directed Information (DI)-based learning. The main difference between these approaches relies on how the privacy term is measured during the training process. The releaser in the CAL method, disposing of supervision from the actual values of the private variables and feedback from the adversary performance, tries to minimize the adversary log-likelihood. On the other hand, the releaser in the DI approach completely relies on the feedback received from the adversary and is optimized to maximize its uncertainty. The performance of these two algorithms is evaluated empirically using real-world SMs data, considering an attacker with access to SI (e.g., the day of the week) that tries to infer the occupancy status from the released SMs data. The results show that, although they perform similarly when the attacker does not exploit the SI, in general, the CAL method is less sensitive to the inclusion of SI. However, in both cases, privacy levels are significantly affected, particularly when multiple sources of SI are included.
\end{abstract}


%
\IEEEpeerreviewmaketitle

\section{Introduction}

Smart meters (SMs) provide advanced monitoring features for power grids by collecting the user's power consumption and reporting them to the utility providers almost in real time. Although this enables several important applications for smart grids (e.g., energy theft prevention, power quality monitoring, demand response, among others~\cite{wang2019}), it also violates the user's privacy, which is a major concern for its wide deployment and adoption \cite{giaconi2018privacy}. Actually, it has been shown that a potential attacker with access to the SMs data can infer sensitive information about the users such as the occupancy status \cite{feng2020deep} or the types of appliances being used by consumers \cite{molina2010}.

In recent years, several studies on privacy-preserving approaches for SMs data sharing were conducted, which can be classified into two main families. On the one hand, the methods in the first family \cite{kalogridis2010privacy,yao2013privacy,tan2013increasing,gomez2014smart,zhang2016cost,giaconi2017smart,li2018information,erdemir2019privacy,giaconi2017optimal,sun2017smart, shateri2020, shateri2020privacycost} use physical resources such as rechargeable batteries, electric vehicles, heating, ventilation, and air conditioning units, etc., to shape the consumed power so that the SMs measurements reveal minimum information about the user's actual power consumption pattern. The methods in the second family \cite{efthymiou2010smart,sankar2013smart, yang2016evaluation, shateri2020privacy, barbarosa2016,tripathy2019privacy, shateri2019deep, shateri2020a} manipulate the SMs data, to be reported to the utility provider, by distorting it in order to prevent the inference of sensitive information by potential attackers, while preserving the usefulness of the data. The study of this paper is focused around this latter family of privacy-preserving methods. We consider the real-time privacy problem, in which an attacker tries to infer sensitive information in an online fashion \cite{shateri2019deep,shateri2020a}, i.e., without  accessing non-causally to SMs data.

Although many of these methods showed good performance on dealing with attackers that have access to the distorted or shaped SMs data, none of them studied the effect of side information (SI), which occurs when an attacker uses additional information to improve its performance. To the best of our knowledge, the only work where SI was considered is~\cite{salehkalaibar2017hypothesis}, in which it is used by the utility provider to predict if the consumers are shaping the power consumption or reporting the actual consumed power. The goal of our study is different. We study the impact of SI at the privacy level, as measured by the accuracy of an attacker trying to guess a sensitive attribute, that is obtained with two different privacy-preserving strategies. First, we consider a causal adversarial learning (CAL) algorithm, which is a generalized version of the privacy-preserving adversarial network (PPAN) approach introduced in~\cite{tripathy2019privacy}, by including the temporal structure of SM data. The CAL algorithm is compared with a recently proposed state-of-the-art method, referred to as the directed information (DI) based learning \cite{shateri2019deep,shateri2020a}, considering attackers with and without SI and using a real world SMs dataset. 





The rest of the paper is organized as follows. In Section \ref{sec:Prob_formul}, the problem formulation of the SMs privacy-utility considering SI is developed for both CAL and DI approaches. Using the proposed formulations, the loss functions for the releaser and adversary are given in Section \ref{sec:model_AL_DI} along with the general training algorithm. Empirical results for both methods are presented and discussed in Section \ref{sec:results}. Finally, some concluding remarks close the paper in Section \ref{sec:conclusion}.

\subsection*{Notation and conventions}
In this paper, a sequence of random variables, or a time series, of length $T$ is shown by $X^T = (X_1, \ldots, X_T)$ while $x^T = (x_1, x_2, \ldots, x_T)$ is used as a realization of $X^T$. The $i^{\text{th}}$ sample in a minibatch used for training of the model is shown by $x^{(i)T} = (x^{(i)}_1, x^{(i)}_2, \ldots, x^{(i)}_T)$. $p_X$ is used to denote the probability distribution function of random variable $X$ and its expectation is shown as $\E[X]$. The Shannon entropy of random variable $X$ is represented by $H(X)$ and KL$(p_X\|  q_X)$ is used to denote the Kullback-Leibler divergence between two probability distributions $p_X$ and $q_X$. In addition, the mutual information (MI) between random variables $X$ and $Y$ is represented as $I(X;Y)$ and ${X \mkv Y \mkv Z}$ denotes a Markov chain among random variables $X$, $Y$ and $Z$. 

\section{Problem Formulation}\label{sec:Prob_formul}

\begin{figure}[htbp!]
	\centering
	\includegraphics[width=0.85\linewidth]{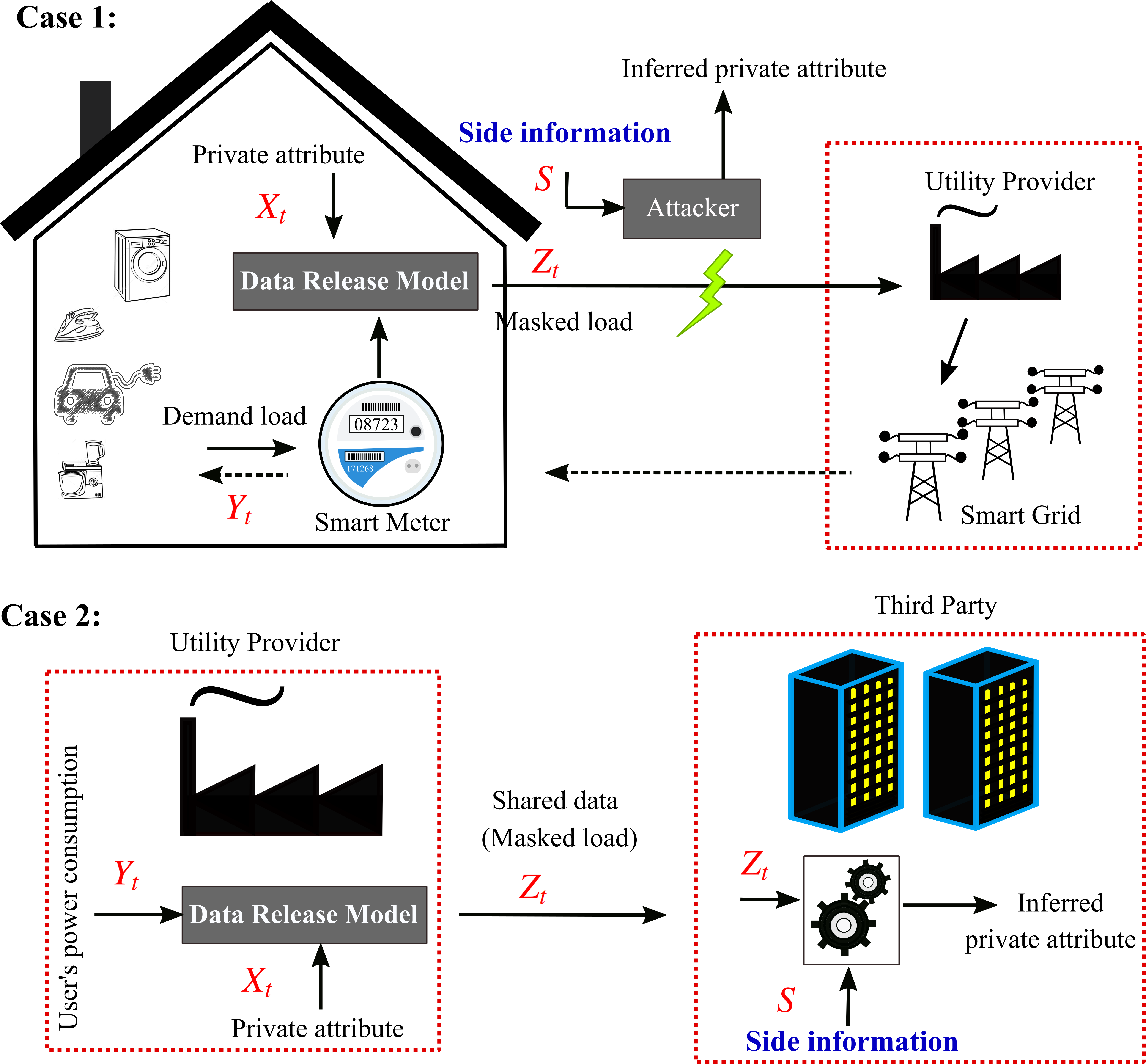}
	\caption{Privacy-preserving SM framework. Case 1: SMs data are communicated with the utility provider. Case 2: Utility provider is sharing the SMs data with a third-party. }
	\label{fig:SMs-privacy-setting}
\end{figure}

Consider a user's power consumption or demand load measured by the SM over $T$ time slots, represented as the time series $Y^T = \{Y_t\}_{t=1}^T$. In addition, let $X^T= \{X_t\}_{t=1}^T$ represent sensitive user information, which could represent power profile details, the presence of individuals at home, the appliances' state (either on or off), etc. As shown in Fig. \ref{fig:SMs-privacy-setting}, the power consumption measurements need to be sanitized before being communicated with the utility provider or shared with a third-party for different data analysis tasks. Concretely, the goal of a privacy-preserving system is to generate a masked load $Z^T$ (referred to as the released time-series), based on knowledge of $Y^T$ and $X^T$ 
, which is similar to $Y^T$ but provides minimum information about the sensitive attribute $X^T$. Note that a potential attacker will try to infer the sensitive information based on the released data. We also assume that the attacker has access to SI (e.g., the day of the week), represented as $S$, which can be used as auxiliary data to improve its inference performance.

In the following subsections, we present two possible formulations of the privacy-preserving problem in this context.


\subsection{Causal adversarial learning (CAL) approach} \label{sec:CAL}

The first approach considered in this study for formulating the privacy-preserving problem is referred to as the CAL method. This approach is a generalized version of the PPAN model introduced in \cite{tripathy2019privacy} but, in addition, includes the temporal correlation and causality involved in the time series data processing. In this setting, the privacy measure is defined as the conditional MI between the released time-series $Z^T$ and sensitive attribute $X^T$, conditioned on the SI $S$. The conditional MI $I(X^T;Z^T|S)$ quantifies the amount of information shared between the released $Z^T$ and sensitive attribute $X^T$ when the side information $S$ is revealed (the reader is referred to \cite{cover2006elements} for details on information-theoretic concepts and properties). The releaser aims to minimize this privacy measure by adding a controlled amount of distortion, say $\varepsilon >0$, to the useful data $Y^T$ so as to keep a desired utility level. Therefore, the problem of finding the optimal releaser can be cast as follows: 
\begin{align} \label{eqtr1} \underset{p_{Z^T|X^T,Y^T}}{\text{min}}  \;  \frac{1}{T} I\left(X^T;Z^T|S\right) \quad \text{s.t.} \quad D(Z^T,Y^T) \le \varepsilon. \end{align}
where $D(Z^T,Y^T)$ is the normalized expected distortion between $Z^T$ and $Y^T$, defined as
\begin{equation} \label{Distortion} D(Z^T,Y^T) \coloneqq \frac{\E[d(Z^T,Y^T)]}{T}, \end{equation}
with $d : \R^T \times \R^T \to \R$ any distortion function.

We now consider an arbitrary conditional distribution $q_{X^T|Z^T,S}$ and note that the conditional MI in \eqref{eqtr1} satisfies the following relation:
\begin{align} \label{var_MI}
I\left(X^T;Z^T|S\right) &= H(X^T|S) - H(X^T|Z^T,S)\nonumber\\
& \overset{\text{(i)}}{=} H(X^T|S) - \sum_{t=1}^{T} H\left(X_t|X^{t-1},Z^T,S\right)\nonumber\\
&\overset{\text{(ii)}}{=} H(X^T|S) + \sum_{t=1}^{T}\bigg[ \E[\log q_{X_t|X^{t-1},Z^T,S}]\nonumber\\
&+ \text{KL}\left(p_{X_t|X^{t-1},Z^T,S}\|q_{X_t|X^{t-1},Z^T,S}\right)\bigg]\nonumber\\
&\overset{\text{(iii)}}{\geq} H(X^T|S) + \sum_{t=1}^{T} \E[\log q_{X_t|X^{t-1},Z^T,S}],\nonumber\\
\end{align}
where (i) is due to the chain rule of entropy, (ii) is a consequence of the definition of the Kullback-Leibler (KL) divergence, and (iii) is due to the fact that KL$(\cdot||\cdot)$ is non-negative. In addition, the expectations are taken with respect to the distribution $p_{X_t|X^{t-1},Z^T,S}$ at each time slot $t$. The equality in \eqref{var_MI} happens when KL$(p_{X^T|Z^T,S}||q_{X^T|Z^T,S})=0$, i.e., when $q_{X^T|Z^T,S}=p_{X^T|Z^T,S}$ (almost surely). Therefore, 
\begin{align} \label{eq:MImax}
I(X^T;Z^T|S) &= H(X^T|S)\quad +\\
&  \underset{q_{X^T|Z^T,S}}{\text{max}} \sum_{t=1}^{T} \E[\log q_{X_t|X^{t-1},Z^T,S}]. \nonumber
\end{align}

Substituting equation \eqref{eq:MImax} in \eqref{eqtr1}, dropping the constant term $H(X^T|S)$, and imposing causality constraints, the privacy-preserving optimization \eqref{eqtr1} can be written as follows:

\begin{align} \label{eq:minmax} \underset{p_{Z^T|X^T,Y^T}}{\text{min}}  \; \underset{q_{X^T|Z^T,S}}{\text{max}} \; & \frac{1}{T} \sum_{t=1}^{T} \E[\log q_{X_t|X^{t-1},Z^t,S}] \\ &\text{s.t.} \quad D(Z^T,Y^T) \le \varepsilon. \nonumber \end{align}


The minmax problem \eqref{eq:minmax} can be interpreted in an adversarial learning context in which the adversary network uses the released data $Z^T$ and $S$ to estimate the posterior $q_{X^T|Z^T,S}$ by maximizing the log-likelihood $\mathbb{E}[\log q_{X_t|X^{t-1},Z^t,S}]$ at time slot $t$, while the releaser attempts to prevent that by minimizing the same quantity. Details for implementing this approach will be presented in Section \ref{sec:model_AL_DI}.

\subsection{Directed information (DI) based approach} \label{sec:DI}

The second methodology considered in this study is the DI-based learning approach, which was introduced in~\cite{shateri2019deep}. In the DI method, by considering the SI, the privacy measure is defined by $I(X^T\rightarrow \hat{X}^T|S)$, the conditional DI between the sensitive attribute $X^T$ and the approximated sensitive attribute $\hat{X}^T$, conditioned on the SI $S$. In \cite{shateri2020a}, it was shown that such a privacy measure can effectively be used to limit the performance of any potential attacker. In this case, the problem of finding the optimal releaser is formulated as the following optimization problem:
\begin{align} \label{eq:DI_optimization} \underset{p_{Z^T|X^T,Y^T}}{\text{min}}  \;  \frac{1}{T} I\left(X^T\rightarrow \hat{X}^T \bigg| S\right) \quad \text{s.t.} \quad D(Z^T,Y^T) \le \varepsilon. \end{align}

Proceeding similarly as in \cite{shateri2020a}, the conditional DI privacy measure can be upper bounded as follows:

\begin{align} \label{eq:di_upper_bound}
I\left(X^T\rightarrow \hat{X}^T \bigg| S\right) \leq T \log|\mathcal{X}| - \sum_{t=1}^{T}H(\hat{X}_t|Z^t,S).
\end{align}


The main advantage of the upper bound \eqref{eq:di_upper_bound}, as discussed in \cite{shateri2019deep}, is its computational tractability for the learning process. By substituting \eqref{eq:di_upper_bound} in \eqref{eq:DI_optimization}, and dropping the constant term $T\log|\mathcal{X}|$, we obtain the following relaxation of the optimization problem:
\begin{align} \label{eq:DI_M} \underset{p_{Z^T|X^T,Y^T}}{\text{min}}  \;  -\frac{1}{T} \sum_{t=1}^{T} H\big(\hat{X}_t|Z^t,S\big) \quad \text{s.t.} \quad D(Z^T,Y^T) \le \varepsilon. \end{align}
Finally, to complete the specification of this problem, we need to define the (optimal) adversary, which is given by the solution to the following optimization problem:
\begin{align} \label{eq:attacker_optimization_problem} & \underset{p_{\hat{X}^T|Z^T,S}}{\text{min}} \;  \text{KL}\big (p_{X^T|Z^T,S}\|p_{\hat{X}^T|Z^T,S}\big ),\quad \text{i.e.}, \nonumber \\ &   \underset{p_{\hat{X}^T|Z^T,S}}{\text{min}} \; - \sum_{t=1}^T \E[\log p_{\hat{X}_t|\hat{X}^{t-1},Z^t,S}], 
\end{align}
where the expectation is taken with respect to $p_{X^t,Z^t,S}$ for each $t$. It should be noted that this is in fact equivalent to the max part in the CAL optimization problem \eqref{eq:minmax} by considering the correspondence $p_{\hat{X}^T|Z^T,S} \sim q_{X^T|Z^T,S}$.


\section{Privacy-Preserving Model}\label{sec:model_AL_DI}

\begin{figure}[htbp!]
	\centering
	\includegraphics[width=0.99\linewidth]{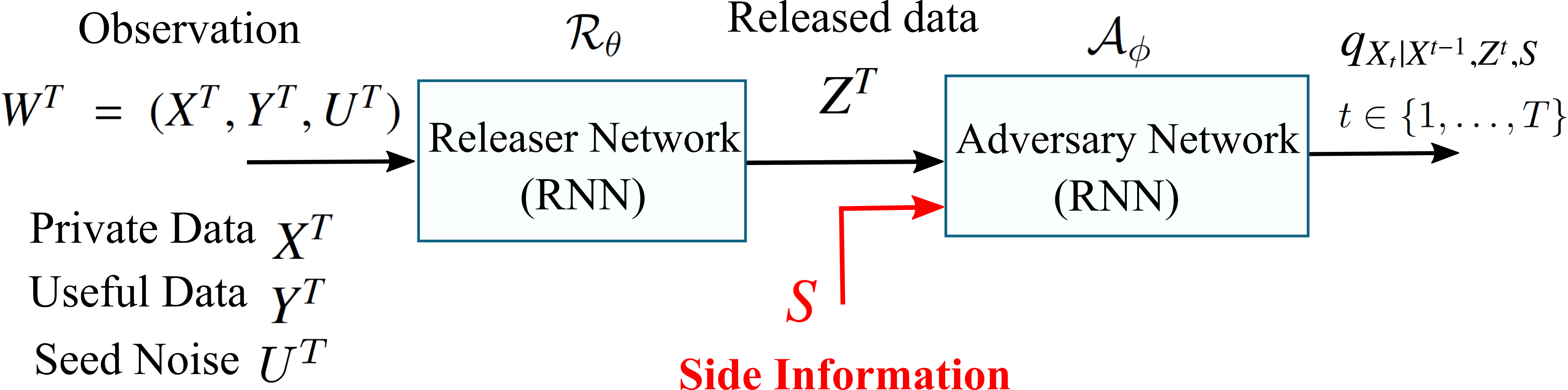}
	\caption{Privacy-preserving framework based on adversarial learning. The observed data $W^T$ is defined as the combination of private and useful data and the seed noise. The seed noise $U^T$ is generated from independent and identically distributed (i.i.d.) samples according to a uniform distribution: $U_t \sim U[0,1]$. }
	\label{fig:SMs-Adversarial}
\end{figure}

In this section, the releaser design problem formulations are tackled by using an adversarial learning framework, as shown in Fig.~\ref{fig:SMs-Adversarial}. In order to exploit the time structure of the data, both the releaser and adversary are modeled using recurrent neural networks (RNNs). RNNs are a class of artificial neural networks specialized for processing sequential data. However, conventional RNNs suffer from the so-called vanishing gradient issue, which leads to difficulties in the training process \cite{bengio1994learning}. To address this problem, the gated RNNs based on long short-term memory (LSTM) cells were introduced in \cite{hochreiter1997long, gers1999learning}, which are widely used in practice currently. For more details on RNNs and LSTMs, the reader is referred to \cite{goodfellow2016}.

The loss functions required for the training of the releaser and adversary networks can be readily defined based on the problem formulations \eqref{eq:minmax}, \eqref{eq:DI_M} and \eqref{eq:attacker_optimization_problem}. On the one hand, for both the CAL and DI models, the adversary network loss function can be written as follows:
\begin{equation} \label{eq:CAL_attacker_loss} \mathcal{L}_{\mathcal{A}}(\phi) \coloneqq \frac{1}{T} \sum_{t=1}^{T} \E\left[-\log q_{X_t|X^{t-1},Z^t,S} \right], \end{equation}
where $\phi$ are the parameters of the adversary network and we consider again the correspondence $p_{\hat{X}^T|Z^T,S} \sim q_{X^T|Z^T,S}$. On the other hand, the loss functions for training the releasers are
\begin{equation} \label{eq:CAL_releaser_loss} \mathcal{L}^{\text{CAL}}_{\mathcal{R}}(\theta, \phi,\lambda) \coloneqq  D(Z^T,Y^T) + \frac{\lambda}{T}\sum_{t=1}^{T} \E\left[\log q_{X_t|X^{t-1},Z^t,S} \right], \end{equation}
\begin{equation} \label{eq:DI_releaser_loss} \mathcal{L}^{\text{DI}}_{\mathcal{R}}(\theta, \phi,\lambda) \coloneqq  D(Z^T,Y^T) - \frac{\lambda}{T}\sum_{t=1}^{T} H\big(\hat{X}_t|Z^t,S\big), \end{equation}
where $\theta$ are the parameters of the releaser network and $\lambda$ controls the privacy-utility trade-off. 

The training process for both cases is presented in detail in Algorithm \ref{AL_pp}.

\begin{algorithm}
    \footnotesize
    \algsetup{linenosize=\tiny}
	\caption{Privacy-preserving data release with side information. Batch size $B$, number of steps to apply to the Adversary $k$, seed noise dimension $m$, and $\ell_2$ regularization parameter $\beta$ are hyperparameters.}
	\label{AL_pp}
	\begin{algorithmic}[1]
	\FOR {number of training iterations}
	    \FOR {$k$ steps}
		\STATE Sample minibatch of $B$ examples $\{ w^{(b)T} = (x^{(b)T},y^{(b)T},u^{(b)T})\}_{b=1}^B$ and generate releases $\{ z^{(b)T}\}_{b=1}^B$.
		\STATE Compute the gradient of $\mathcal{L}_{\mathcal{A}}(\phi)$ with respect to $\phi$, empirically approximated with the minibatch data.
		\STATE Update $\phi$ by applying the RMSprop optimizer \cite{hinton2012neural}.
		\ENDFOR
		
		\STATE Sample minibatch of $B$ examples $\{ w^{(b)T} = (x^{(b)T},y^{(b)T},u^{(b)T})\}_{b=1}^B$ and generate releases $\{ z^{(b)T}\}_{b=1}^B$.
		
		\STATE Compute the gradient of $\mathcal{L}_{\mathcal{R}}(\theta,\phi,\lambda)$ with respect to $\theta$, empirically approximated with the minibatch data.
		\STATE Use $\textrm{Ridge}(L_2)$ regularization \cite{hastie2005elements} with value $\beta$ and update $\theta$ by applying RMSprop optimizer.
		\ENDFOR
	\end{algorithmic}
\end{algorithm}

\section{Results and Discussion} \label{sec:results}

\subsection{Dataset and model parameters}

In this work, we use the electricity consumption and occupancy (ECO) dataset published by \cite{beckel2014eco}, which includes 1 Hz electricity usage measured by SMs along with the occupancy labels of five houses in Switzerland. For the sake of simplicity, the dataset is re-sampled every one hour and samples over one day ($T=24$) are considered. The dataset is split into training and test sets with a ratio of \mbox{85:15}, while $10 \%$ of training data is considered as the validation dataset. Using the validation dataset, the values of the hyperparameters (including batch size $B$, number of adversary training step $k$, seed noise dimension $m$, and regularization value $\beta$) were tuned to achieve the best privacy-utility trade-off. Both the CAL and DI models are trained to hide the occupancy labels by distorting the electricity consumption. The performance of these models are evaluated based on the performance of an attacker, trained in a supervised manner, which attempts to infer the occupancy labels. Three different cases are considered:
\begin{itemize}
	\item Case 1: No SI is considered for training the privacy-preserving model nor for training the attacker.
	\item Case 2: The day of the week associated to the SMs samples is used as SI for both training the privacy-preserving model and the attacker.
	\item Case 3: The day of the week and the month of the year are used as SI for both training the privacy-preserving model and the attacker.
\end{itemize}

 
The structures of the releaser, adversary, and attacker are similar for both the CAL and DI methods. The releaser is composed of 4 LSTM layers (each including 64 cells) with $\beta = 1.5$, $k=4$, $m=8$, and $B=128$, while the attacker is made of 3 LSTM layers (each including 32 cells). For the second and third cases, the adversary network consists of 3 LSTM layers (each including 32 cells), while for the first case it is composed by 2 LSTM layers (each including 32 cells).

\begin{figure*}[t]
	\centering
	\includegraphics[width=0.8\linewidth]{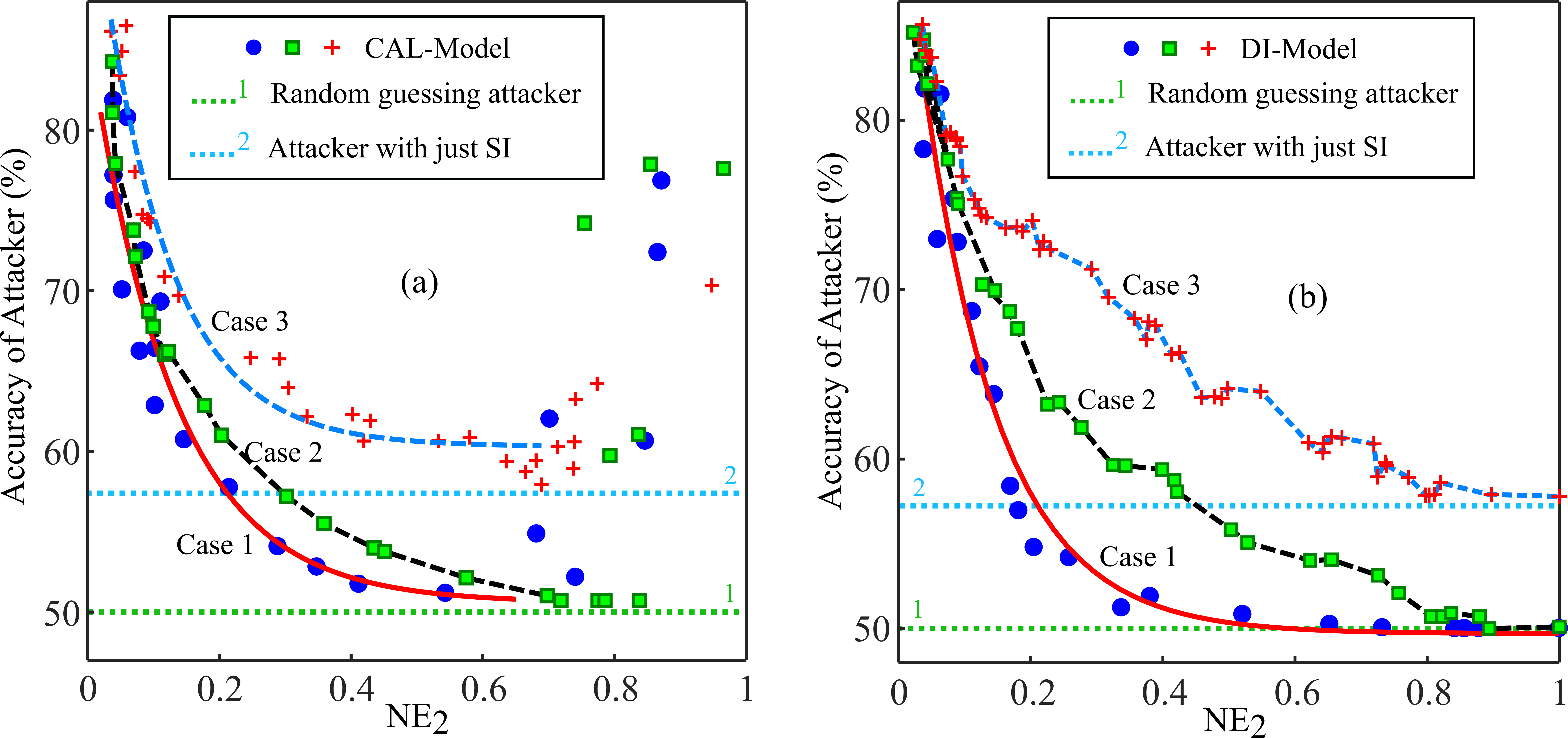}
	\caption{(a) CAL model, (b) DI model, Privacy-utility trade-off for house occupancy inference. Case 1: no side information; Case 2: day of the week as side information; Case 3: day of the week joined with  month of the year as side information.}
	\label{fig:CAL-DI-tradeoff}
\end{figure*}

To evaluate the amount of distortion added by the releaser network, we use the Normalized Error (NE) measure, defined as follows:
\begin{equation}\label{eq:NE2} \text{NE}_2 \coloneqq \frac{\E\left[ \| Y^T - Z^T \|_2 \right]}{\E\left[ \| Y^T \|_2 \right]}. \end{equation}
In addition, the performance of the attacker in inferring the sensitive data is measured using the balanced accuracy~\cite{mosley2013balanced}, presented below:
\begin{equation}\label{eq:BA} \text{Balanced Accuracy} \coloneqq \frac{1}{2} \left( \frac{c_{11}}{c_{11}+c_{12}} + \frac{c_{22}}{c_{22}+c_{21}} \right), \end{equation}
where $c_{ij}$ represent the fraction of examples of class $i$ classified as class $j$. Using these metrics, the privacy-utility trade-off for the CAL and DI models are presented in Figs. \ref{fig:CAL-DI-tradeoff}.

\begin{figure}[htbp!]
	\centering
	\includegraphics[width=.8\linewidth]{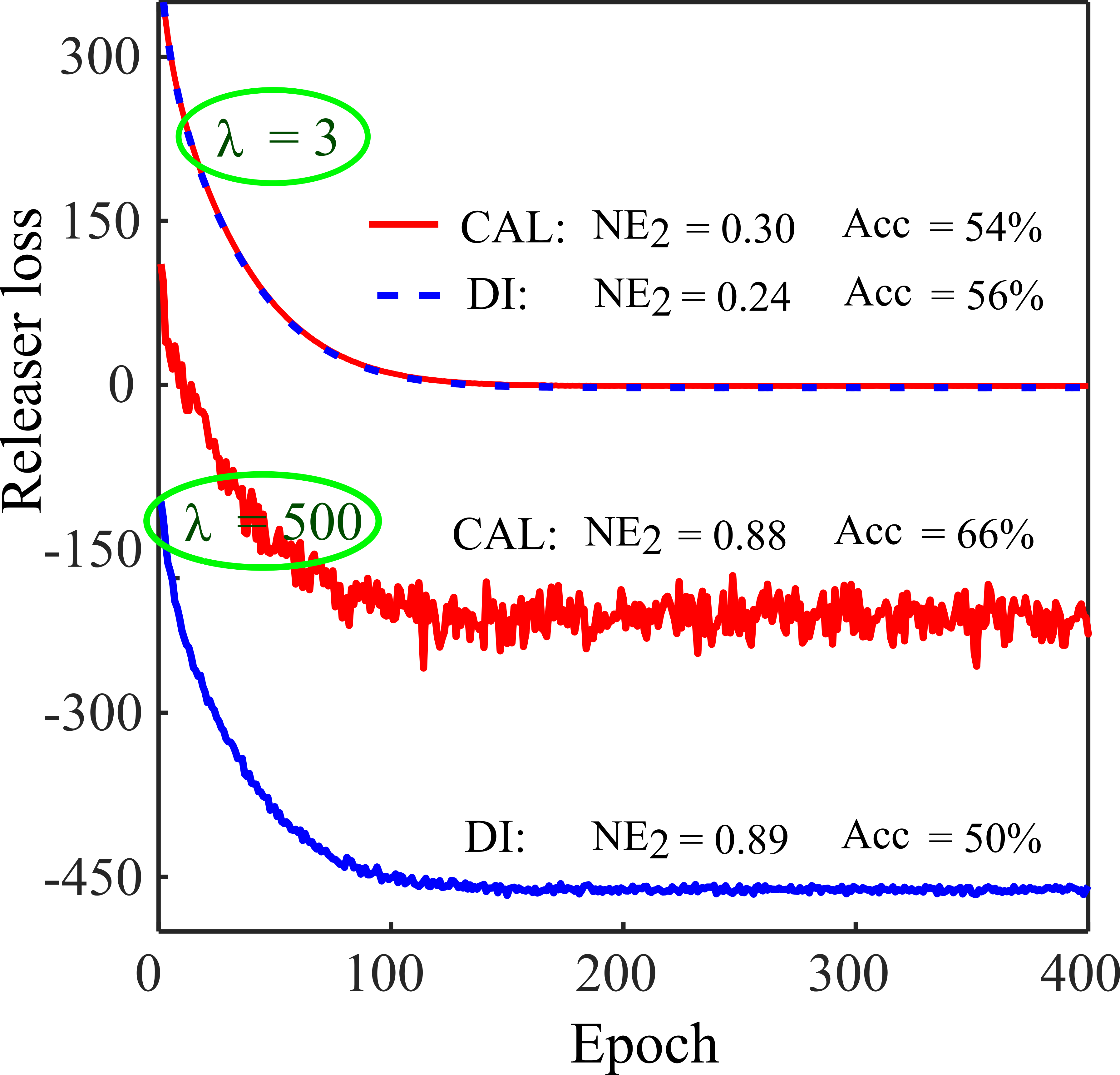}
	\caption{Examples of the releaser network loss function versus training epochs for both DI and CAL models.}
	\label{fig:BothLoss}
\end{figure}

Considering the first case, in which no SI is taken into account, it can be seen that both models have very similar performances. However, the CAL model behaves erratically for large values of distortion, in the sense that the accuracy of the attacker does not monotonically decrease as the distortion increases. From a practical perspective, however, the high distortion region for which this happens is not particularly interesting, since the utility of the distorted data is very low in such cases. To understand this issue, we studied the evolution of the releaser loss function during training for both models, considering different values of $\lambda$. As can be seen in Fig.\ref{fig:BothLoss}, for small values of $\lambda$, the releaser loss functions behave very similarly with a clean convergence pattern. However, as $\lambda$ increases and more distortion is allowed, the convergence of the loss for the CAL model becomes noisier than for the DI model. In other words, for large values of $\lambda$ (i.e., the full privacy regime) the DI model, by maximizing the conditional entropy, can push the adversary towards a random guessing classifier, while the CAL model, by maximizing the cross-entropy, does not seem to work well. However, in the area where there is a balance between privacy and distortion, the cross-entropy seems to be effective in controlling the privacy-utility trade-off. 




For the other cases, where SI is included in the models, it is clearly seen that the privacy performance is degraded, which is expected since the attacker has more information about the sensitive attribute to perform the inference task. It should be noted that the baseline for full privacy changes for the different cases. Indeed, an attacker trained and tested with just SI for estimating the sensitive attribute suggests a balanced accuracy of 50.7\% and 57.8\% as the baseline for Cases 2 and 3, respectively. In particular, for the Case 3 in which both the day of the week and month of the year are considered, the attacker performance is improved in a very significant way. In fact, the models can not completely fool the attacker even when arbitrarily large distortion is allowed. This phenomenon can be understood because the SI provides some prior information for the attacker, which can be exploited independently from the amount of noise added to the power measurements by the releaser (see attacker performance with just SI in Fig. \ref{fig:CAL-DI-tradeoff}. On the other hand, the CAL model seems to be less sensitive (in the low distortion range) to SI than the DI model. This can be justified by revising the loss functions for each model (see \eqref{eq:CAL_releaser_loss} and \eqref{eq:DI_releaser_loss}). The releaser in the DI method completely relies on the adversary uncertainty and therefore is unsupervised, while the CAL releaser gets supervision from the actual sensitive labels, which can make it more effective. 


\begin{figure}[htbp!]
	\centering
	\includegraphics[width=0.8\linewidth]{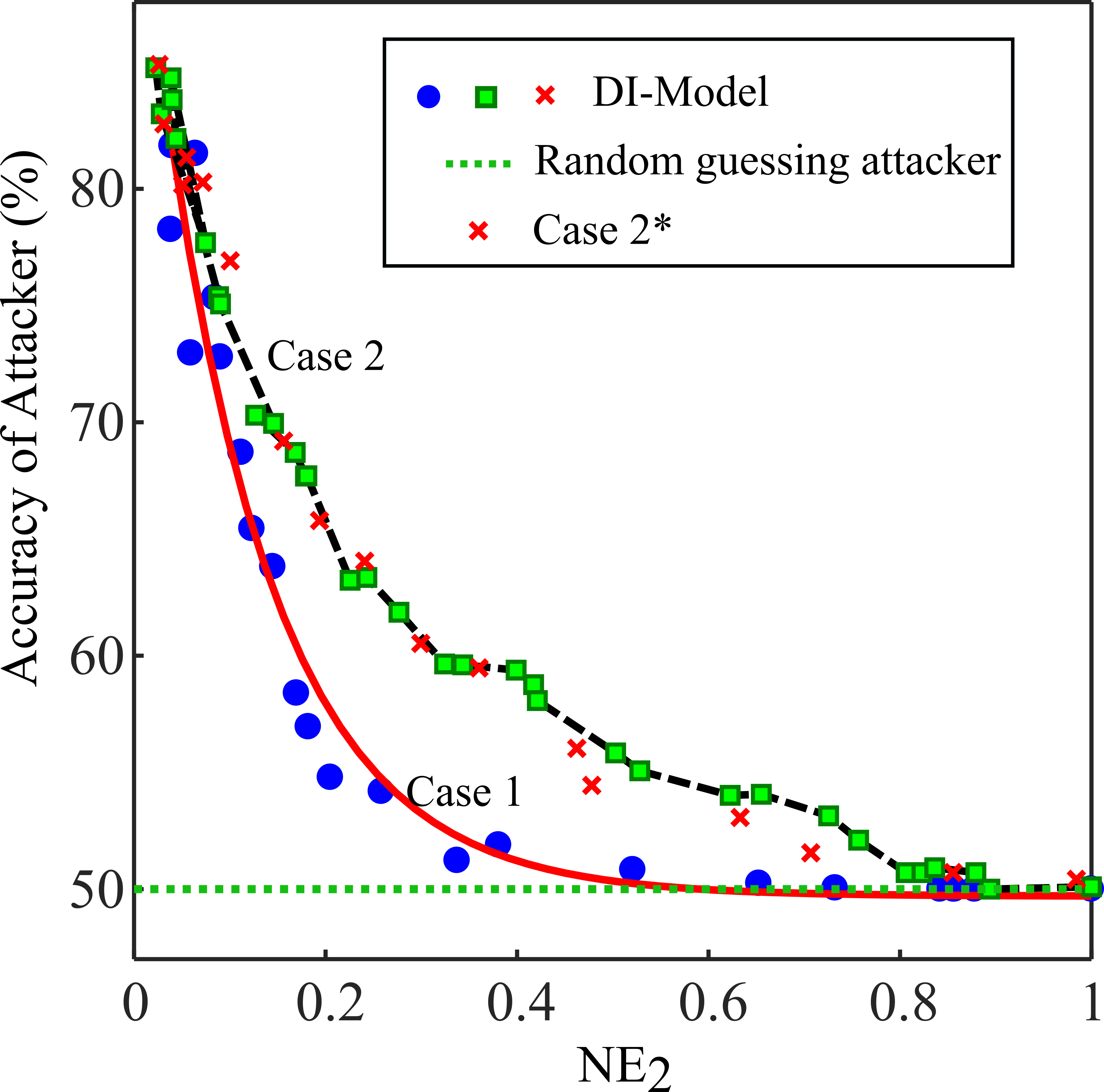}
	\caption{DI model: Privacy-utility trade-off for house occupancy inference. Case 2$^{*}$ refers to Case 2 but including side information also at the input of the releaser.}
	\label{fig:SI_REL}
\end{figure}

A final experiment that we conducted was including the SI at the input of the releaser network for Case 2, which means that the privacy-preserving mechanism can change its behavior according to the day of the week. As can be seen in Fig.\ref{fig:SI_REL}, however, this is not helpful in reducing the gap between Case 2 and Case 1 in Fig. \ref{fig:CAL-DI-tradeoff}. This might be due to the fact that the SI $S$ can be readily inferred from $X^T$ and $Y^T$. For example, for the data set in our study the day of the week can be predicted from the $X^T$ and $Y^T$ with balanced accuracy of more than 85\%. This suggests that SI is in some sense a redundant input for the releaser network.  

\section{Concluding Remarks} \label{sec:conclusion}

In this work, we took into account the effect of SI (correlated with sensitive information) on the formulation of SMs privacy-preserving mechanisms. Concretely, two distortion-based real-time privacy-preserving models were presented and implemented using a deep adversarial learning framework. The privacy-utility trade-offs associated with both models were then investigated and compared. For the case in which SI is not considered, both models perform very similarly, except for large distortion values, where the CAL model showed instability issues. For the other cases, in which the attacker had access to SI, the privacy levels are significantly affected, but the CAL model was shown to be more robust than the DI model to the inclusion of SI. The privacy degradation was particularly noticeable when multiple sources of SI were considered jointly. This result clearly shows how it is possible to overestimate the privacy level attained by a privacy-preserving mechanism  ignoring sources of SI, a particularly serious hurdle for offering actual privacy guarantees. This observation raises several questions for future research: How should privacy be evaluated when not all possible sources of SI can be taken into account? How can we effectively model the attacker prior information in general?





\section*{Acknowledgment}
This work was supported by Hydro-Quebec, the Natural Sciences and Engineering Research Council of Canada, and McGill University in the framework of the NSERC/Hydro-Quebec Industrial Research Chair in Interactive Information Infrastructure for the Power Grid (IRCPJ406021-14). This project has received funding from the European Union’s Horizon 2020 research and innovation programme under the Marie Skłodowska-Curie grant agreement No 792464.

\bibliographystyle{ieeetr}
\bibliography{main}

\end{document}